\documentclass[12pt]{article}
\pdfoutput=1
\usepackage[centertags]{amsmath}
\usepackage{amsfonts} \usepackage{amssymb} \usepackage{amsthm}
\usepackage{graphicx}
\usepackage{pstricks}
\usepackage{pgf,pgfarrows,pgfnodes}
\usepackage{bbm}
\usepackage{bm}

      \let\g=\gamma   
\let\e=\epsilon

\let\G=\Gamma

\newcommand{\be}{\begin{equation}}
\newcommand{\ee}{\end{equation}}
\newcommand{\bea}{\begin{eqnarray}}
\newcommand{\eea}{\end{eqnarray}}
\newcommand{\ba}{\begin{array}}
\newcommand{\ea}{\end{array}}
\def\nn{\nonumber}

\newcommand{\ft}[2]{{\textstyle\frac{#1}{#2}}}
\newcommand{\Z}{{\mathbb Z}}
\newcommand{\R}{{\mathbb R}}

\begin{document}
\begin{titlepage}
\begin{flushright}
{ROM2F/2011/03}\\
\end{flushright}
\begin{center}
{\large \sc Gauge theories on $\Omega$-backgrounds  from \\
non commutative  Seiberg-Witten curves }\\
\vspace{1.0cm}
{\bf F.Fucito}, {\bf J. F. Morales},  {\bf D. Ricci Pacifici}\\
 I.N.F.N. Sezione di Roma Tor Vergata\\ and \\{\sl Dipartimento di Fisica, Universit\'a di Roma ``Tor Vergata''\\
Via della Ricerca Scientifica, 00133 Roma, Italy}\\
and\\
{\bf R.Poghossian}\\
Yerevan Physics Institute,\\
Alikhanian Br.2, 0036 Yerevan, Armenia
\end{center}
\vskip 2.0cm
\begin{center}
{\large \bf Abstract}\\\end{center}

We study  the dynamics of a ${\cal N}=2$ supersymmetric $SU(N)$ gauge theory with fundamental or adjoint matter in presence of  a non trivial  
$\Omega$-background along a two dimensional plane. The prepotential and chiral correlators of the gauge theory 
can be obtained, via a saddle point analysis, from an equation which can be viewed  as a non commutative version of the ``standard'' 
Seiberg and Witten curve.  
 
 \vfill
\end{titlepage}

\tableofcontents

\section{Introduction and Summary }

 In \cite{Seiberg:1994rs,Seiberg:1994aj}, Seiberg and Witten (SW) provided us with a 
beautiful solution  for the quantum prepotential of a four-dimensional Yang-Mills gauge theory  with ${\cal N}=2$ supersymmetry 
(to be called ${\cal N}=2$ gauge theory from now on)
at the non perturbative level. In this set up, the dynamics of the
gauge theory is encoded in  the periods of a two-dimensional Riemann surface defined by a holomorphic curve.  
 
With the introduction of localization techniques  \cite{Nekrasov:2002qd,Flume:2002az,BFMT} in the study of non perturbative effects for such teories, 
the emergence of the curve and of the prepotential  was directly derived from 
multi-instanton computations \cite{Nekrasov:2003rj}. Localization requires a deformation of the spacetime geometry,
the so called $\Omega_{\epsilon_1,\epsilon_2}$ background,  
regularizing the spacetime volume and leading to a finite multi instanton partition function $Z(\epsilon_1,\epsilon_2,q)$.  
The prepotential of the ${\cal N}=2$  is identified with the free energy   ${\cal F}=-\epsilon_1 \epsilon_2 \, \ln Z $ in the limit
$\epsilon_{1,2}\to 0$. The SW curves emerge 
from a saddle point evaluation of the multi instanton partition function in the limit $\epsilon_\ell \to 0$  \cite{Nekrasov:2003rj}.  
Besides its regularization role, the parameters $\epsilon_\ell$
 can be interpreted in string theory as the vacuum expectation values of certain Ramond-Ramond fields \cite{Billo:2006jm}. 
Moreover the  coefficients in the double $\epsilon$ expansion  of ${\cal F}$ have been related \cite{Antoniadis:2010iq} to the so called ${\cal F}_{g,n}$ 
topological string amplitudes first computed in \cite{Morales:1996bp}.

      It is natural to ask how the SW theory is modified in the presence of the $\Omega_{\epsilon_1,\epsilon_2}$ backgrounds.  
Here we address this question focusing on the simplest case $\epsilon_\ell=(0,\epsilon)$.  
This type of background has recently received a lot of attention 
due to its relations with  quantum integrable systems \cite{Nekrasov:2009rc}. 
 
The limit $\epsilon_1\to 0$ is also interesting from the point of view of the so called AGT
correspondence \cite{Alday:2009aq} since it corresponds to the quasiclassical limit
 where the central charge  of the underlying CFT becomes very large. 
     
The case of the $SU(N)$  theory with fundamental matter has been  very recently studied in \cite{Poghossian:2010pn} 
where an `` $\epsilon$ deformed SW curve" was derived from a saddle point analysis of the instanton partition function.    
The chiral correlators of the SYM theory were computed in terms of the integrals of $\lambda_J=x^J d z$ where 
$z(x)$ is a holomorphic function determined by the  saddle point equations which, 
in the limit $\epsilon_\ell \to 0$, lead to the ``standard'' SW curve.    
In this paper we revisit these results and extend them to the case with adjoint matter where the 
$\epsilon$ deformed differential is found as a solution of an integral equation.

We rely on a saddle point analysis and encode the information about the saddle point solution
into a single holomorphic function $z(x)$\footnote{In the main text we  work
with the variable $w(x)$ connected to $z(x)$ via $w(x)=e^{-z(x)}$.}.
This  function specifies the instanton distribution dominating the partition function in the limit $\epsilon_1\to 0$.
   It is useful to think of gauge instantons as D(-1) branes bounded to D3 and D7 branes.  
After localization, the instantons distribute along the plane transverse to both the D3 and D7 branes at distances of 
order $\epsilon_\ell$ from the positions of the D3 branes.
 In the limit $\epsilon_\ell\to 0$ they condense into continuous intervals centered at the D3 brane positions.
One can think of this configuration as a two dimensional electrostatic system
 made out of metallic plates of charge -2 near D3 branes and point like charges +1 at the 
 D7-brane positions\footnote{See Appendix B and expecially (\ref{elecpro}) for a justification of this statement.}. The  
saddle point equations  then become the conditions that the potential
 is constant along the metallic plates and the  imaginary part of the holomorphic function $z(x)$ is identified with 
 the electrostatic potential. Finding the SW curve is then equivalent to solve the electrostatic problem.  
 Such techniques have also appeared in connection with matrix models in \cite{Kazakov:1998ji} and \cite{Hoppe} for an earlier reference.
  The case of finite $\epsilon$,   can be thought of as a discretization of this electrostatic problem where the metallic plates 
split into  infinite number of 
 dipoles  with dipole length $\epsilon$. The saddle point equation once again is expressed as 
 the condition that the potential at the center of any  dipole coincide. 
  
   Alternatively the saddle point equations can be written as functional equations for  $z(x)$. 
 We will show that these equations in the case of fundamental matter can be thought  of as a ``non commutative" or ``quantum" version of  the SW curve.   
Evidences of such 
non commutative structure in the case of adjoint matter will be also   presented.  Our result provides a further support to the 
proposal in \cite{Nekrasov:2009rc} for a relation  of the gauge dynamics in the $\epsilon$ background to Toda and Calogero-Moser quantum integrable models.
  
To be concrete, let us consider a SW curve written as
    $$
    W(x,e^z)=0
    $$ 
 and a one form differential $\lambda=x dz$.
    For a $SU(N)$  gauge theory with fundamental matter, $W$ is a polynomial of order $N$ in $x$ and  
order two in $e^z$. 
   We claim that the $\epsilon$ deformed dynamics is encoded into the non commutative
   version of the curve
   \be
   W(\hat x , e^{\hat z} ) | \Psi \rangle =0       \qquad     [\hat z, \hat x]=\epsilon   \label{nc}
   \ee
    This non commutative relation  can be  realized by taking  either $z=\epsilon \partial_x$
    or $x=-\epsilon \partial_z$. In the first case $| \Psi \rangle$ is realized as a function
    $\Psi(x)$. (\ref{nc}) becomes a difference  equation relating $\Psi(x)$ to $e^{\pm \hat z} \Psi(x)=\Psi(x\pm \epsilon)$.
 Specifying to the case of fundamental matter we will show how this equation reproduces
the one following from a saddle point analysis of the multi instanton partition function.
 In particular the deformed SW differential will be related to $\Psi(x)$ in a simple way. The difference equation will be solved and 
written in a continuous fraction form that will give
   the full $\epsilon$ dependence at each order in $q$. This  generalizes a similar result in \cite{Poghossian:2010pn} were the   
$U(1)$ solution was written in terms of  hypergeometric functions.
  In the case of adjoint matter, $W$ is given in terms of  theta functions and the analysis
  is more involved, since
the order in $e^z$ grows along with the  $q$ expansion. The resulting difference equation can be still
solved order by order in $q$ and we tested it against a direct multi instanton computation.

Alternatively (\ref{nc}) can be seen as a differential equation of order $N$
for $\tilde{\Psi}(z)=| \Psi \rangle$ after the identification $\hat{x} =-\epsilon \partial_z$. This was the point of view taken in
\cite{Mironov:2009uv,Mironov:2009dv,Mironov:2009ib} for the pure gauge theory and \cite{Maruyoshi:2010iu} for the case of  adjoint matter.
 The SW differential was identified with $\lambda=d\ln \tilde{\Psi}(z)$ and
the periods where checked against the formulae for the leading $\epsilon$ corrections. In the case of an $SU(2)$ gauge theory, 
this leads \cite{Maruyoshi:2010iu,Alba:2009ya,Mironov:2010ym,Mironov:2010pi} to very robust tests of the 
correspondences between $\epsilon$ deformed  gauge theories  and CFT's or quantum integrable models.
These results inspired our proposal. We remark that in this formulation the resulting $N$ order differential equation can be  
typically solved only perturbatively in $\epsilon$,  in contrast with the  difference equation which determines the full $\epsilon$ 
dependence of the differential. Clearly $\Psi(x)$ and $\tilde{\Psi}(z)$ are related to each other via a Fourier transform.

 It would be nice to explore the implications of this non commutative structure 
 in the M theory and type IIA descriptions of Seiberg-Witten theory.  
In  IIA theory \cite{Witten:1997sc}, the gauge theory is realized in terms of N D4 branes suspended between two NS5 branes. 
The endpoints of the D4 branes on the two NS5 branes behave as charges in an appropriate sense consistently with our electrostatic analogy. 
This picture lifts to M theory, where the brane system is replaced by a single M5 brane wrapping the two dimensional curve and the four dimensional 
spacetime. 
The eleventh dimensional circle is identified with the imaginary part of $z$ which is compact due to the trivial identification  $z \sim z+2\pi i$. 
The holomorphicity of $z(x)$ ensures that the imaginary part of  $z(x)$ admits the interpretation of a two dimensional electrostatic potential 
as claimed. Our results suggest that the $\epsilon$ deformation can be realized in these pictures
by promoting the spacetime coordinates $x,z$ to non commutative variables.

The paper is organized as follows:  in section 2 we review the computation of the instanton partition
function. In section 3 we use a saddle point analysis to evaluate such partition function and
derive the general form of the saddle point equations, prepotential  and chiral correlators 
of the  gauge theory. In section 4 and 5 we specify the results to the case of $SU(N)$ gauge
theories with fundamental and adjoint matter respectively. In the Appendices we collect several
useful technical data. In appendix A we discuss some properties of the counting function 
encoding the information about the saddle point solution. In Appendix B we present an alternative
derivation of the saddle point equations based on the extremization over the so called profile function describing the  shape of the  
Young tableaux which give the leading contribution to the
instanton partition function. In Appendix C we present some tests of the $\epsilon$ deformed 
SW differentials against direct multi instanton computations. In Appendix D we comment on the
connection to quantum integrable systems.

\section{Instanton partition functions}

In this section we review the computation of the instanton partition
functions and of the chiral correlators for a ${\cal N}=2$  gauge theory
with gauge group $SU(N)$ and matter in the fundamental or adjoint
representation.

We start from ${\cal N}=4$  with gauge group $SU(N)$ realized in terms of open strings 
connecting N D3 branes in flat ten dimensional spacetime. After a mass deformation this becomes the so called ${\cal N}=2^*$ theory, i.e. 
${\cal N}=2$ gauge theory with a massive adjoint hypermultiplet.   In the limit of large mass it reduces to a pure ${\cal N}=2$ gauge theory. 
Alternatively, 
the adjoint matter can be projected out via an orbifold projection, i.e by modding out an internal four dimensional space by a discrete group. 
Fundamental matter can be added by including certain number of D7 branes. 

The brane picture describes also the non perturbative objects of the gauge theory.  
Instantons  of winding number $k$ are realized by introducing $k$ D(-1) branes. The instanton moduli are associated to the massless modes of 
those open strings which have at least one end on a D(-1) brane. The prepotential and chiral correlators of the gauge theory  are computed  
by integrals over the resulting moduli space. 
The explicit evaluation of these integrals can be carried out with the help of localization
techniques that reduce the integrations to the evaluation of a determinant at a finite number of
isolated fixed points of the moduli space symmetries. To achieve complete localization, both
gauge and Lorentz symmetries should be broken. Gauge symmetries can be broken by turning
on a vacuum expectation value for the adjoint scalar field $\langle \Phi \rangle ={\rm diag} \{ a_u \} $
in the SU(N) vector multiplet. Lorentz symmetries can be broken by turning on a non trivial 
background $\Omega_{\epsilon_1,\epsilon_2}$ on the four dimensional spacetime.
The parameters $\{ a_u,\epsilon_{1,2} \}$ parametrize the Cartan subgroup of the
 gauge and Lorentz symmetries, $SU(N)\times SO(4)$, broken by the background.
The flat space result can be recovered by sending $\epsilon_{1,2}\to 0$ at the end of the computation.

The parameters $a_u, \, u=1,\ldots,N$ we introduced before specify the positions of
the D3 branes along a transverse  plane. 
The positions of the instantons along this plane
will be denoted by $\phi_I$, $I=1,...k$. The $\Omega_{\epsilon_1,\epsilon_2}$ background creates a non trivial potential which penalizes 
those instantons trying to move away from the D3 brane and thus regularizing the spacetime volume ${\rm vol}_{\R^4}\sim {1\over
\epsilon_1 \epsilon_2}$. As a result, instantons distribute near
the points $a_u$'s where the D3 branes sit and the $\phi_I$'s can be accommodated in a Young
tableau centered at $a_u$ with boxes of sizes $\epsilon_{1,2}$.
The instanton partition function can then be written as
\bea
Z_{\rm inst}(q) &=& \sum_{k=0}^\infty q^k Z_k =1+\sum_{k=1}^\infty q^k \, \int   {1\over k!  }   \prod_{I=1}^k \frac{d{\phi_I}}{2\pi i}~ z_k(\phi)\nn\\
&=& 1+\sum_{k=1}^\infty {q^k\over k!} \int \prod_{I=1}^k
 {d\phi_I  \over 2\pi i} \,  \prod_{I,J}^k {}'  
 D(\phi_{I}-\phi_{J}) \prod_{I=1}^k  Q_0(\phi_I)\nn\\
&=&\sum_{\vec Y}     \prod_{I,J}^k {}^\prime  D(\phi^Y_{I}-\phi^Y_{J}) \prod_{I=1}^k 
\,  q \,Q_0(\phi^Y_I)
\label{zzz}
\eea
with $D(x)$, $Q_0(x)$   the contributions  of open
strings with both or a single end respectively on the D(-1) instantons. 
The prime in the product denotes the omission  for $I=J$  
given by the replacement $D(0) \to D'(0)$. 
The form of these functions depends on the specific matter content and will be given below in this section.

The integrals in (\ref{zzz}) are evaluated by closing the contours in the complex plane
and picking up the corresponding residues\footnote{We take the pole prescription
${\rm Im} \epsilon_1 >> {\rm Im} \epsilon_2 >0$ and close the contours in the upper half plane.}.
The relevant poles of the integrands (those whose contribution do not sum up to zero) (see (\ref{fund}),(\ref{adj}) below)
are specified by the $N$ Young tableaux set $\{ Y_u \}$ with total number of boxes $k$. Explicitly
\be
 \phi_I^Y=\phi_{u,i_u,i_u'}=a_u+(i_u'-1)\epsilon_1+(i_u-1) \epsilon_2  \label{phiin}
 \ee
 with $i_u',i_u$ running over the rows and columns of the Young tableau $Y_u$. 
The instanton partition function is then given by summing over all possible Young tableaux.

The partition function encodes the information about the gauge theory
prepotential via the identification
\be {\cal 
F}(\epsilon_1,\epsilon_2,q)=\sum_{k=1}^\infty {\cal F}_k q^k =-\epsilon_1 \epsilon_2 \ln
Z(q)
\ee
with $Z(q)=Z_{\rm pert}(\tau) Z_{\rm inst}(q)$ given in terms of the perturbative $Z_{\rm pert}(\tau)$
(tree level and one loop) and the instanton contribution $Z_{\rm inst}(q)$ given by (\ref{zzz}). 
The details of $Z_{\rm pert}(\tau)$ will not be relevant to our analysis since $Z_{\rm pert}(\tau)$ does not depend on the details of the instanton
configuration dominating the saddle point. The classical information of where the Young tableaux are located will be supplemented later
by giving the periods, $a_u$, of the SW curve.
An explicit form of $Z_{\rm pert}$ can be found in Appendix \ref{profileappend}.
In the limit $\epsilon_{\ell}\to 0$,  ${\cal F}$ reduces to the
SW prepotential of the gauge theory.

Besides the prepotential, the chiral dynamics is completely specified by
the correlators of the adjoint chiral field $\Phi$. They are computed by integrals of the same type we have just described with extra $\phi_I$ insertions.
A generating function for all chiral correlators can be  written as
\bea
\langle {\rm tr}\, e^{z \Phi} \rangle
=\sum_u e^{z a_u}- Z^{-1}_{\rm inst}(q)\sum_{k=1}^\infty {q^k \over k!} \,
\int    \prod_{I=1}^k \frac{d{\phi_I}}{2\pi i}~
 z_k(\phi)\,  \sum_{J=1}^k e^{z \phi_J} \prod_{l=1}^2(1-e^{z \epsilon_l}) \label{chircor}
\eea
 
 \subsection*{$SU(N)$ plus fundamental matter}

  The functions appearing in (\ref{zzz}) and (\ref{chircor}) in the case of a $SU(N)$ 
  gauge theory with fundamental matter are 
  \bea
   D(x) &=& {x (x+\epsilon_1+\epsilon_2)\over (x+\epsilon_1) (x+\epsilon_2)} \nn\\
 Q_0(x)  &=&   {  M(x) \over   P_0(x+\epsilon_1+\epsilon_2 ) P_0(x ) } \label{fund}
 \eea
 with
 \be
 P_0(x) =\prod_{u=1}^N (x-a_u) \qquad
 M(x) =   \prod_{a=1}^{N_f} (x-m_a)
 \ee
The various contributions
to $D(x)$ come from D(-1)D(-1) strings while those in $Q_0(x)$
come from D(-1)D3 and D(-1)D7 open strings. Each contribution
accounts for one complex moduli. In particular the denominator of
$D(x)$ comes from those moduli describing the position of
the instanton in the four dimensional spacetime. The numerator in
$D(x)$ accounts for the ADHM constraints and the $U(k)$ gauge
redundance. The denominators in $Q_0(x)$ describe the D(-1)D3  moduli. 
Finally the contributions to $M(x)$ come from massless fermionic moduli
in the D(-1)D7 sector with $N_f$ the number of
D7 branes and $m_a$ parametrizing the masses of the corresponding
fundamental matter.

\subsection*{$SU(N)$ plus an Adjoint hypermultiplet}

    The functions appearing in (\ref{zzz}) and (\ref{chircor}) in the case of $SU(N)$
  gauge theory with adjoint matter read
  \bea
   D(x) &=& {x (x+\epsilon_1+\epsilon_2) (x+m+\epsilon_1) (x+m+\epsilon_2)\over (x+\epsilon_1) (x+\epsilon_2)(x+m) (x+m+\epsilon_1+\epsilon_2)} \nn\\
 Q_0(x)  &=&   {  P_0(x-m ) P_0(x+m+\epsilon_1+\epsilon_2 )  \over   P_0(x ) P_0(x+\epsilon_1+\epsilon_2 ) }
 \label{adj}
 \eea
 with $m$ the mass of the adjoint hypermultiplet. Now the denominator of $D(x)$ describes the
 position of the instantons in an eight dimensional space and the numerators are the generalized ADHM
 constraints. Similarly the D(-1)D3 open strings contain, besides the standard moduli connected to the instanton radius and orientations, 
a set of auxiliary fields contributing to the numerator of $Q_0(x)$.

\section{Saddle point analysis}

 In this section  we use a saddle point technique to determine 
the instanton partition function in the limit $\epsilon_1\to 0$ with $\epsilon_2=\epsilon$ finite. 
We rederive here some results of \cite{Poghossian:2010pn} in a form more suitable for various generalizations.   
For simplicity we take  $a_u,\epsilon_{1,2}$ to be real.
Our formulae will be later extended to the complex plane by supplementing $\epsilon_{1,2}$
  with a small and positive imaginary part.
Exponentiating the products in (\ref{zzz}) one can write the partition function in the form
 \bea
Z_{\rm inst} &=&\sum_{\vec Y}   {\rm e}^{    \sum_{I,J} {}^\prime \ln D(\phi^Y_I-\phi^Y_J)+\sum_I \ln(q Q_0(\phi^Y_I))}   \label{zr}
 \eea
 It is convenient to introduce the density function
  \bea
 \rho(x) &=&\epsilon_1 \sum_I \delta(x-\phi_{I} )
\label{rho0}
 \eea
describing the distribution of the instantons along the real line.
Moreover in the limit $\epsilon_1 <<  x$ the function $D(x)\approx 1$ as can be seen from formulae (\ref{fund}) or (\ref{adj}) above.  We write
\be 
e^{\ln D(x)} \approx e^{\epsilon_1
G(x)}
\ee 
 Plugging this  into (\ref{zr}) and using
(\ref{rho0}) one finds
 \be
Z_{\rm inst}=
 \int D\rho\, {\rm e}^{ {1\over \epsilon_1} {\cal H}_{\rm inst}(\rho)}   \label{zr2}
 \ee
 with
 \be
\framebox[1.15\width ][c]{${\cal H}_{\rm inst}(\rho) = {1\over 2} \int_{\R\times \R} dx dy \rho(x) \rho(y) G_s
(x-y)+\int_{\R} dx \rho(x) \ln \left[ q Q_0(x)   \right]$}  \label{hhh}
\ee
and
 \be
  G_s(x)= G(x)+G(-x) =\lim_{\epsilon_1 \to 0} {1\over \epsilon_1} \ln D(x)D(-x)
   \label{gg}
 \ee
twice the even part of $G(x)$. The main contribution to the partition function will come from
instanton configurations $\rho(x)$ extremizing (\ref{hhh}). We remark that the $Z_{\rm pert}$ we introduced earlier
does not depend on the instanton density $\rho(x)$ and therefore it is irrelevant for the discussion of the saddle point.

\subsection{The  density function}

The integral in (\ref{zr}) runs over the density functions of type (\ref{rho0}) with $\phi^Y $ specified
by the corresponding Young tableaux set. According to (\ref{phiin}), in the limit $\epsilon_1\to 0$,
the instantons form a continuous distribution starting at
\be
x_{ui}^0=a_u +(i-1) \epsilon   \qquad u=1,\ldots N \quad i=1,\ldots \infty \label{xui0}
\ee
and ending at some $x_{ui}$, given by the top end of the $i^{th}$ column in $Y_u$.
 This implies in particular that the sequence $x_{ui}$  decreases when $i$ grows (keeping
 fixed u) reaching $x_{ui}=x_{ui}^0$ at some $i$ where the Young tableau ends.
The instanton position $\phi_I^Y$ can then be described in terms of the continuous variable
 \be
 \phi_I^Y=\phi_{u,i} \in [x_{ui}^0,x_{ui} ]
 \ee
The sums over I can then be written as
\be
\sum_I  ={1\over \epsilon_1} \sum_{ui} \int_{x_{ui}^0}^{x_{ui}} d\phi_{ui}
\ee
 and the density function (\ref{rho0}) becomes
 \footnote{Here we use $\int_a^b \delta(x-y) dy=-\theta(x-y)\Big|_a^b$ with $\theta(x)$ the
  Heaviside step function.}
  \bea
 \rho(x) &=&
  \sum_{ui} \left[\theta(x-x^0_{ui}) -\theta(x-x_{ui})\right]
= \left\{
  \begin{array}{cc}
  1 & x \in   [x_{ui}^0,x_{ui} ]\\
  0 & {\rm otherwise}
\end{array}\right.
\label{rho}
 \eea
  It is important to stress that the set $\{ x_{ui} \}$ completely specifies the Young tableaux
set and therefore the density function $\rho(x)$. In particular the empty Young tableaux set
corresponds to taking $x_{ui}=x_{ui}^0$, i.e. $\rho(x)=0$.
In Figure 1 we display (in green) the instanton distribution along the $\phi$ line
associated to a  given Young tableau profile.  
 The columns of this tableau are to be thought of as  made out of a large number of thin boxes 
 ending at the points $x_{ui}$  determined by the saddle point equations. 

\begin{figure}[t!]
\label{ftableaux}
\begin{center}
\includegraphics[scale=0.4]{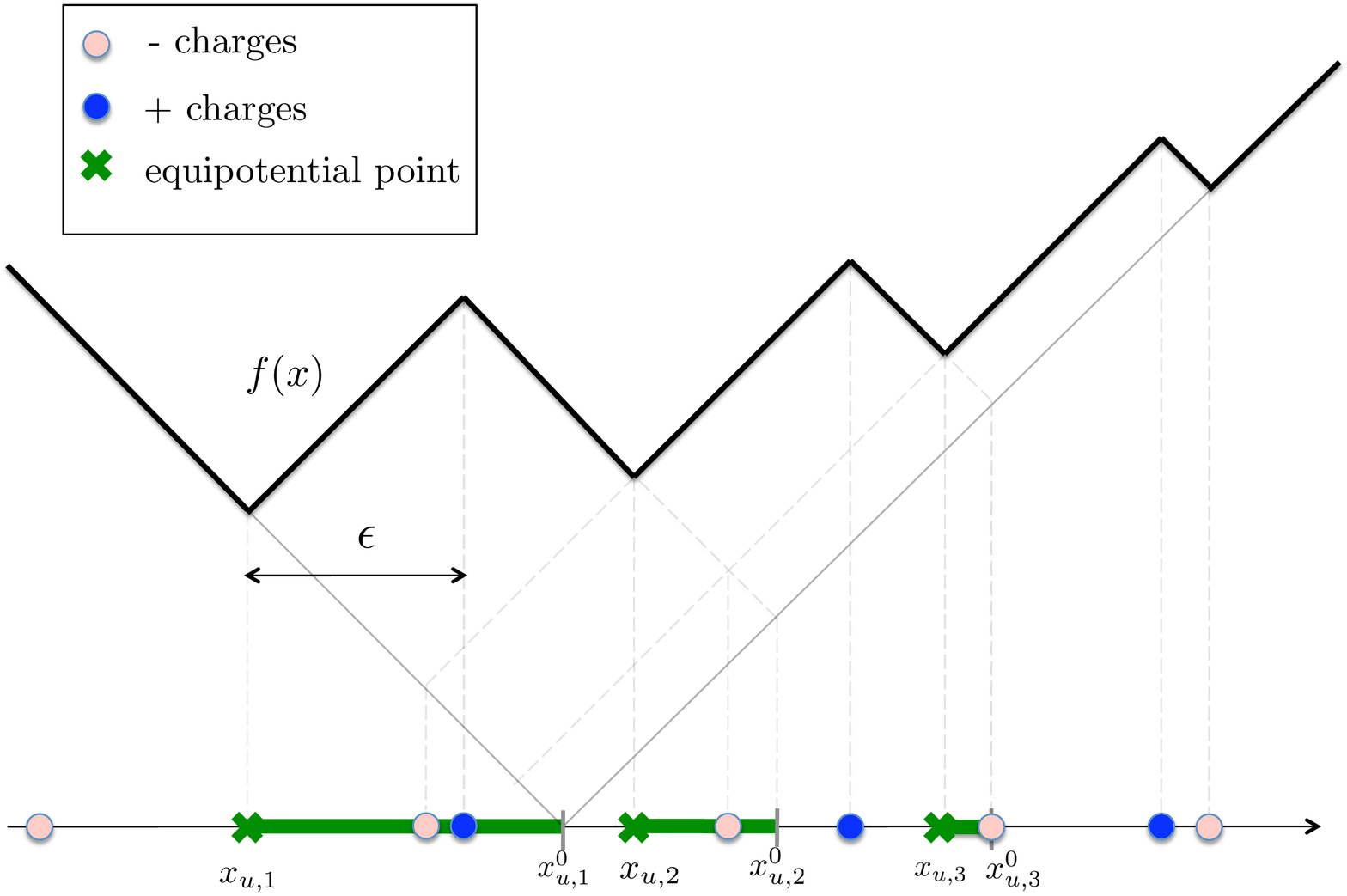}
\label{fig: tableaux}
\caption{ \footnotesize Instanton distribution associated to a Young tableau centered at $a_u$. We display
in green the intervals of non trivial instanton density, i.e.  $\rho(x)=1$. By blu and red bullets
we indicate plus and minus charges in the auxiliary electrostatic problem where crosses 
stand for points of constant potential.}
\end{center}
\end{figure}

\subsection{Saddle point equations}

 The leading contribution to the integral (\ref{zr}) in the limit $\epsilon_1\to 0$ comes then
from a configuration $\{ x_{ui} \}$ extremizing (\ref{hhh})
\be
 \framebox[1.15\width ][c]{${\delta {\cal H}_{\rm inst}(\rho) \over \delta
x_{ui}} =\int_{\R} dy \rho(y) G_s (x_{ui}-y)+ \ln \left[ q Q_0(x_{ui} )   \right]
=0 $} \label{saddle0}
\ee
The function $G_s(x)$ defined by (\ref{gg}) can be conveniently written in the form
(see formulae (\ref{funddata}) and (\ref{adjdata}) below)
 \be
G_s(x)=\sum_{a=1}  (-)^a{d\over dx} \ln (x+\alpha_a) \label{gsgen}
\ee 
for some $\alpha_a$ that depends on the specific matter content.  

The integral in the first term in (\ref{saddle0})  becomes
 \be  
 \int_{\R} dy \rho(y) G_s (x-y)= 
\sum_{u,i}\int_{x^0_{ui}}^{x_{ui}}dy\, G_s (x-y)
=\sum_{a=1} (-)^{a+1} \ln {\cal Y}(x+\alpha_a)
\label{intformula} 
\ee 
with  ${\cal Y}(x)$ given by 
 \be
 {\cal Y}(x) =  \prod_{v=1}^N \prod_{i=1}^\infty \left({ x- x_{vi}  \over
  x- x^0_{vi}  } \right) \label{yyyy}
\ee
 We notice that the function ${\cal Y}(x)$ has zeros at $x_{ui}$  and poles at $x_{ui}^0$.
The convergence of the infinite product entering  in (\ref{yyyy}) follows from the
fact that $x_{ui}=x_{ui}^0$ for $i$ large enough. 

 It will be also convenient to define the $x_{ui}^0$ independent ratio
\bea
      w(x)&=&{{\cal Y}(x-\epsilon)\over {\cal Y}(x) P_0(x)
      }=  \lim_{L\to \infty} {1\over (-L \epsilon )^N}
 \prod_{u=1}^N \prod_{i=1}^L {x-x_{ui}-\epsilon\over x-x_{ui}}
\label{yw}
   \eea
The right hand side follows after writing the products over $i$ in the definitions  of ${\cal Y},{\cal Y}_0$ up to a cut off $i=L$ 
and sending $L$ to infinity (see Appendix \ref{stoolkit} for an alternative derivation and details).
 Notice that all $x_{ui}^0$ dependence cancel out in this limit. 
(\ref{yw}) shows that $w(x)$ has zeros at $x_{ui}+\epsilon$ and poles at $x_{ui}$.

For large $x$ the functions ${\cal Y}(x)$ and $w(x)$ behave as
 \be
 {\cal Y}(x) \approx  1  \qquad w(x) \approx  {1\over x^N }  \label{wxlarge}
 \ee
The functions ${\cal Y}(x)$ or $w(x)$ encode the information about the saddle point
 configuration  solving (\ref{saddle0}) and therefore the saddle point equation can be viewed
 as a functional equation for one of these functions. In particular, we will see how the
functional equation for $w(x)$ provides an $\epsilon$ deformed version of the familiar SW curves
encoding the corrections induced by the $\Omega_{\epsilon_1,\epsilon_2}$ background on the gauge theory.

 \subsection{Chiral correlators}

The saddle point analysis in the last section can be also applied to the integrals (\ref{chircor})
defining the chiral correlators. Notice first that the two integrals share the same saddle point since
they only differ by $\phi_I$ insertions. Denoting the
saddle point instanton configuration and the instanton number, giving the leading contribution in the sum
over $k$, by $\{ \phi_{I,{\rm ext}} \}$ and $k_{\rm ext}$,  one finds\footnote{At leading order $Z_{\rm inst}\approx q^{k_{\rm ext}}\, Z_{k_{\rm ext}}$ and 
there is a cancellation between the numerator and the denominator. }
 \be
\langle {\rm tr}\, e^{z \Phi} \rangle
 \approx \sum_{u=1}^N e^{z a_u}- (1-e^{z \epsilon_1}) (1-e^{z \epsilon_2})
\sum_{i=1}^{k_{\rm ext}} e^{z \phi_{i,{\rm ext} }  }
 \ee
Keeping only the leading contribution in the $\epsilon_1\to 0$ limit
and expanding in powers of $z$ both sides of this equation, one finds
\bea
\langle {\rm tr} \Phi^J \rangle & \approx & \sum_{u=1}^N  a_u^J +
 \int_{\R}    dy\,\rho(y)  {d\over dy}  \left[y^J-(y+\epsilon)^J\right]
     \nn\\
     & \approx & \sum_{u=1}^N  a_u^J +
  \sum_{u=1}^N \sum_{i=1}^\infty \left[x_{ui}^J -(x_{ui}+\epsilon)^J+(x^0_{ui}+\epsilon)^J-x^{0~J}_{ui}\right]    \label{xuij}
 \eea
 where we use the equivalent writings (\ref{rho0}) and (\ref{rho}) of $\rho(x)$ 
 in the first and second lines respectively. 
(\ref{xuij})  is just what one would have obtained from the results in \cite{Losev:2003py,Flume:2004rp} after performing the $\epsilon_1\to 0$ limit.
The sums on the r.h.s. can be written as a contour integral as
\bea \langle {\rm tr} \Phi^J \rangle
 &=&  \int_{\gamma} {dy\over 2\pi i}  \,y^J\,   \partial_y \ln
\left[  {P_0(y ) {\cal Y}(y) \over  {\cal Y} (y-\epsilon)}
\right]\label{thphij0}  \nn\\
&=&
  -
\int_{\gamma} {dy\over 2\pi i} \,y^J\, \partial_y \ln w(y)
\label{trphij}
\eea
where $\gamma$ is a counterclockwise contour surrounding the whole real line.
In writing this we use the fact that ${\cal Y}(x)$ and $P_0(x)$ have zeros of order one at $x_{ui}$ and $a_u$.
Moreover, ${\cal Y}(x)$ has poles at $x_{ui}^0$ which 
cancel against the zeroes of $P_0(x)$ leading to the $x_{ui}^0$ independent function, $w(x)$ in the second line of (\ref{thphij0}). 
The integral on the right hand side picks up the
residues at all these poles. Multiplying by $x^{-J-1}$ this equation (with $x$ a point outside of $\gamma$), summing
over $J$ and thinking of $\gamma$ as a contour integral around the single pole at $y=x$ 
one can write the
generating function of the chiral correlators as
\be
\framebox[1.15\width ][c]{$\langle \,{\rm tr} {1\over x-\Phi}
\rangle=-\partial_x \ln w(x)$}\label{xfi}
\ee
On the other hand the
prepotential of the gauge theory
 is computed using
 \bea
 q {d
{\cal F}\over d q} &=& -\epsilon   \left( q { \partial {\cal H}_{\rm inst} \over \partial x_{ui}
}\,{\partial x_{ui}\over \partial q} +  q {\partial   {\cal H}_{\rm inst}    \over \partial q}    \right)
  =  -\epsilon \int_{\R} \rho(y) dy \nn\\
  &=&  -\epsilon \sum_{ui} (x_{ui}-x_{ui}^0)=-k_{ext}\epsilon_1\epsilon
   \label{fdq}
 \eea
where the saddle point equation ${ \partial {\cal H}_{\rm inst} \over \partial x_{ui} }=0$
 has been used. We notice that the prepotential is related to the number of boxes, $k_{ext}$, in the
 Young tableau giving the leading contribution to the partition function.
The expression  on the right hand side can be related to $\langle {\rm tr} \Phi^2\rangle$
using the equation in the second line of (\ref{xuij}) for $J=2$.  One finds
 \be
  \langle {\rm tr} \Phi^2 \rangle =\sum_u a_u^2+2 q {d {\cal F}\over d q}
   \label{qdqf}
  \ee
  the well known Matone relation\cite{Matone:1995rx}.
  The classical vevs are defined by
\be
\framebox[1.15\width ][c]{$a_u =   - \int_{\gamma_u} {dy\over 2\pi i}  \,y\,    \partial_y   \ln
w(y) $}
\label{au}
\ee
 with $\gamma_u$ a contour surrounding all the $x_{ui}$ and $x_{ui}+\epsilon$ for a fixed $u$.

 Summarizing, the prepotential and the chiral correlators of an ${\cal N}=2$  gauge theory
 in presence of a non trivial $\epsilon$ background can be written in terms of the integrals
 of the $\epsilon$ deformed SW differential
 \be
 \lambda= - x\, d \ln w(x)
\ee
 with $w(x)$ encoding the details of the saddle point solution $\{ x_{ui} \}$ of  (\ref{saddle0}). In
 the next sections we will rewrite the saddle point equations as functional equations for $w(x)$
 that reduce to the SW curves in the limit of $\epsilon\to 0$.

\section{ $SU(N)$ plus fundamental matter}

In this section we specify our general formulae to the case of
$SU(N)$ with fundamental matter. The $\epsilon$ deformed SW curve
in this case  was derived
in \cite{Poghossian:2010pn}.  The solution for the U(1) case was found and written in terms
of hypergeometric functions. Here we review these results and present a solution
for the $SU(N)$ case.  
We also show that the saddle point equation can be interpreted as a `` non commutative" version of the 
``standard'' SW curve with $\epsilon$ measuring the scale of non commutativity.

\subsection{Saddle point equations}

~ From (\ref{fund}) and (\ref{gg}) one finds
 \bea
   G_s(x) &=& {d\over dx} \ln \left(x+\epsilon\over x-\epsilon\right)   \nn\\
  Q_0(x)  &=&   {
M(x) \over   P_0(x) P_0(x+\epsilon ) }\label{funddata}
 \eea
Notice that in this case $ \alpha_a$ defined in (\ref{gsgen}) assumes the values 
\be
 \alpha_a=(-\epsilon,\epsilon)    \label{alphafund}
\ee
Plugging this into (\ref{intformula}) and using (\ref{yw}) and (\ref{funddata}) one finds \bea
 \int_{\R} dy\, G_s(x-y)\, \rho(y)
  & =&  \ln {{\cal Y}(x-\epsilon)  \over  {\cal Y}(x+\epsilon) }\nn\\
&=& \ln \left( {w(x) w(x+\epsilon) M(x) \over  Q_0(x)} \right)  \label{intgfund}
 \eea
 The saddle point  equation (\ref{saddle0}) then  becomes
   \bea
   \framebox[1.15\width ][c]{$ 1- q\, M(x_{ui})\,w(x_{ui})\,w(x_{ui}+\epsilon) =0   $}
     \label{sadfund}
   \eea
  This equation can be solved recursively for $x_{ui}$ order by order in
  $q$. Indeed, at order $q^L$, one can write
\be x_{ui}=a_u+(i-1)\epsilon+\sum_{k=i}^L \lambda_{ui;k} \, q^k
\label{xguess} \ee
 and approximate $w(x)$ by $w_L(x)$ given by restricting the
 products over $i$ up to $i=L$ 
 \be
 w(x) \approx w_L(x) ={1\over P_0(x-L \epsilon )}
 \prod_{u=1}^N \prod_{i=1}^L {x-x_{ui}-\epsilon\over x-x_{ui}}
 \label{wwl}
\ee
  Plugging (\ref{xguess}) and (\ref{wwl}) into (\ref{sadfund}) and
  solving for $\lambda_{ui;k}$ one finds the $x_{ui}$ characterizing
  the saddle point solution.
 Here we will follow an alternative route and extract the function $w(x)$ from a functional
 equation (see next section) that generalizes the SW curve to finite $\epsilon$.

We conclude this section with a  comment on the electrostatic interpretation of 
the saddle point equation. To this aim, we notice that  
 (\ref{sadfund}) can be rewritten as
\be
U(x_{ui})= {\rm const}   \label{uconst}
\ee
with
\be
U(x)=\sum_{vj} \ln \left[ { x-x_{vj}-\epsilon\over   x-x_{vj}+\epsilon } \right]
+\sum_{a=1}^{N_f}    \ln (x-m_a)     
\ee
 The function $U(x)$ can be interpreted as the two dimensional electrostatic potential
 generated by a set of positive charges at $x_{ui}+\epsilon$, $m_a$ and negative charges
 at $x_{ui}-\epsilon$. The saddle point  equation, (\ref{uconst}), is nothing else that the condition
 that the potential at all points $x_{ui}$ has the same value. We illustrate in figure 1
 the electrostatic problem for $N_f=0$ and a generic Young tableau diagram with $L=3$ columns.
 Positive and negative  charges are displayed as blue/black and pink/grey bullets.  
 The  net charge is -2, independently of the value of $L$.     

\subsection{Deformed SW differential}

(\ref{sadfund}) can be rewritten as a holomorphic equation in the complex plane
 by  introducing a function $f(x)$ defined as
     \be
f(x)  ={1-q\, M(x-\epsilon) w(x)w(x-\epsilon)   \over  w(x) }
  \ee
   It is easy to see that $f(x)$ has no poles since the zeros of the denominator
   at $x=x_{ui}+\epsilon$ are also zeros of the numerator according
    to the saddle point equation (\ref{sadfund}).
    In addition at large $x$, using the asymptotics (\ref{wxlarge}),
    one finds $f(x)\approx x^N$ for $N_f<2 N_c$ and $f(x)\approx (1-q) x^N$
    for $N_f=2N_c$\footnote{For this choice the theory is conformal and $q$ is dimensionless.}. We can then conclude that function $f(x)$ is a polynomial of order $N$. We write 
    \be
 f(x)=P(x)=\prod_u (x-e_u)
  \label{ffff}\ee
  in the case of $N_f<2 N_c$ and  $f(x)=(1-q) P(x)$ for $N_f=2N_c$.
For $N_f<2 N_c$ the saddle point equation becomes
  \be
   \framebox[1.15\width ][c]{$ q\, M(x-\epsilon) \,w(x)w(x-\epsilon)+w(x) P(x)-1 =0  $}   \label{swdef}
  \ee
For  $N_f=2 N_c$,  equation (\ref{swdef}) still holds after replacing $P(x)\to (1-q)P(x)$. 
 At $\epsilon=0$, equation  (\ref{swdef}) reduce to the familiar SW curve for SU(N) with matter in the fundamental representation.

The meaning of the parameters $e_u$ can be understood considering the limit of large $x$ where $w(x)$
behaves as
\be
w(x)= \sum_{i=0}^\infty  c_i x^{-N-i}  \label{winf}
\ee
Plugging this into  (\ref{swdef}) and solving for the first few $c_i$'s one finds
\be
 -\partial_x \ln w(x)=\sum_{J=0}^{2N-N_f-2}   \sum_{u=1}^N {e_u^J\over  x^{J+1}} +o(x^{-2N+N_f})
\ee
i.e., using (\ref{xfi})
 \be
\langle  {\rm tr} \phi^J  \rangle = \sum_{u=1}^N e_u^J   \qquad J<2N-N_f
\ee
This implies that the parameters $e_u$ specifying the polynomial $P(x)$  can be interpreted as the 
quantum analog of  the v.e.v.'s $a_u$\cite{Dorey:1996ez}.

A solution of the deformed SW (\ref{swdef}) can be easily written
 in a continuous fraction form
\bea
w(x) &=& {1\over P(x)+q\, M(x-\epsilon) \,w(x-\epsilon)} \nn\\
&=& {1\over P(x)+{ q\, M(x-\epsilon) \over P(x-\epsilon)+{q\, M(x-2 \epsilon)  \over P\left(x-2\epsilon \right)+\ldots }}}
\eea
Expanding in powers of $q$ one finds $w(x)=\sum_k w_k(x) q^k$ with
\bea
w_0(x) &=& {1\over P(x)}\nn\\
{w_1(x)\over w_0(x)} &=&  -{ M(x-\epsilon) \over P(x-\epsilon)
P(x)}
\eea
and so on. Plugging this into (\ref{au}) one finds
 \bea
 a_u &=& - \sum_{i=0}^\infty {\rm Res}_{y=e_u+i  \epsilon}\,  y\,\partial_y \ln
w(y)\\
&=& e_u- {q\over P'(e_u)} \left[ {M(e_u- \epsilon)
  \over  P(e_u-\epsilon)}+  {M(e_u)
  \over  P(e_u+\epsilon)}   \right]+O(q^2) \label{aefund}
 \eea
Similarly, chiral correlators and the prepotential can be computed plugging $w(x)$ into (\ref{trphij}) and (\ref{qdqf}).

\subsection{Quantizing the SW curve}

We conclude this section by showing that the  $\epsilon$ deformed
equations just obtained can be derived from  a ``non commutative"  version  of the SW curve.
To this aim, we write the undeformed SW curve for the $SU(N)$  gauge theory
with fundamental matter  in the form
\be
W(x,e^z)=  q\, M(x) \,e^{-z}+P(x) - e^{z}  =0
\label{swdefq}
  \ee
  In these coordinates the SW differential reads $\lambda=x dz$. The variable $z$ is related to the $w$ function entering the  
definition of the deformed SW differential
 via $z=-\ln w$.

  We claim that the $\epsilon$  deformed dynamics can be extracted from the eigenvalue equation
   \be
   W(\hat x , e^{\hat z} ) | \Psi \rangle =0         \label{ncw}
   \ee
that follows from promoting the SW coordinates to non commutative
variables satisfying
\be
     [\hat z, \hat x]=\epsilon   \label{nc2}
 \ee
    Taking  $\hat z=\epsilon \partial_x$, $\hat x=x$ and  $| \Psi \rangle=\Psi(x)$, the eigenvalue
    problem  (\ref{ncw}) reduces to the difference equation
 \be
 \framebox[1.15\width ][c]{$ q\, M(x) \Psi(x-\epsilon)+P(x ) \Psi(x)-\Psi(x+\epsilon) =0 $} \label{ybe}
 \ee
 where we used $e^{\pm \hat z} \Psi(x)=\Psi(x\pm \epsilon)$.
  Dividing by $\Psi(x+\epsilon)$ and identifying
  \be
w(x)={\Psi(x-\epsilon)\over \Psi(x)}
\ee
  one reproduces the deformed SW curve
(\ref{swdef}) after the trivial redefinition of the parameters $e_u$ or equivalently
the replacement $P(x) \to P(x+\epsilon)$.

 It is worth noting the close relation of (\ref{ybe}) 
 to the Baxter's T-Q equation which emerges in the context of 2d integrable models. In fact for the pure  $U(N)$ case 
 with no extra matter ($M(x)\equiv 1$) this difference equation exactly coincides with the Baxter's equation for the 
 periodic $N$ particle Toda chain \cite{Gaudin:1992ci}. We are not aware of similar relations in the case of matter in the fundamental
representation.

\section{$SU(N)$ plus adjoint matter}

In this section we specify to the ${\cal N}=2$ gauge theory with an adjoint hypermultiplet.

\subsection{The saddle point equations}

~ From (\ref{adj}) and (\ref{gg}) one finds
 \bea
     G_s(x) &=& {d\over dx} \ln \left({(x+\epsilon)(x+m)(x-m-\epsilon)
\over (x-\epsilon)(x-m)(x+m+\epsilon)} \right) \nn\\
Q_0(x)  &=& { P_0(x-m ) P_0(x+m+\epsilon ) \over P_0(x+\epsilon )
P_0(x ) } \label{adjdata} \eea Notice that the parameters $\alpha_a$ in (\ref{gsgen})
are given  now by \be
\alpha_a=(-\epsilon,\epsilon,-m,m,m+\epsilon,-m-\epsilon)    \label{alphaadj} \ee Plugging this into
(\ref{intformula}) one finds \bea
  \int_{\R} dy\, G_s(x-y)\, \rho(y)
  &=& \ln  {{\cal Y}(x-\epsilon) {\cal Y}(x-m){\cal Y}(x+m+\epsilon)\over
  {\cal Y}(x+\epsilon) {\cal Y}(x+m){\cal Y}(x-m-\epsilon)  }\nn\\
&=& \ln \left( {w(x) w(x+\epsilon)\over w(x+m+\epsilon) w(x-m)\, Q_0(x)} \right)
\label{intgadj} \eea
 The saddle point equation (\ref{saddle0}) then  becomes
   \bea
    \framebox[1.15\width ][c]{$
    1-q {w(x_{ui})w(x_{ui}+\epsilon)\over    w(x_{ui}+m+\epsilon) w(x_{ui}-m) } =0$} \label{sadadj}
   \eea
(\ref{sadadj}) can be solved again perturbatively for
$x_{ui}$ with the help of   (\ref{xguess}) and (\ref{wwl}).

We notice that the saddle point equations can be reinterpreted again as the
equipotential condition (\ref{uconst}) for the two dimensional electrostatic potential
\be
U(x)=\sum_{vj} \ln \left[ { (x-x_{vj}-\epsilon)(x-x_{vj}-m)
 (x-x_{vj}+m+\epsilon) 
 \over   (x-x_{vj}+\epsilon)   (x-x_{vj}+m) (x-x_{vj}-m-\epsilon)    } \right]
\ee
Now the charges distribute near $a_u$ and $a_u\pm m$. The net charge
 around  $a_u$ is $-2$ as in the case of a pure ${\cal N}=2$ gauge theory (see figure 1).
  Similarly one can see that the net charges around $a_u\pm m$  is $+1$\footnote{ 
  Alternatively one can think of $U(x)$ as the superposition
  $  U(x)=U_m(x)+U_{-m}(x+\epsilon)$ with
$
U_m(x)=\sum_{vj} \ln \left[ { (x-x_{vj}-\epsilon)(x-x_{vj}-m)
 \over  ( x-x_{vj}   ) (x-x_{vj}-m-\epsilon)    } \right]
$.  $U_m(x)$ is the potential generated by a set of plus-minus charges with
net charge -1 and +1 at $a_u$ and $a_u+m$ respectively. }.
 
\subsection{Deformed SW differential}

Like in the case of fundamental matter, matching the $m$ independent zeros and poles
on the right hand side of the saddle point equation (\ref{sadadj}), one can write
     \be
        1-q {w(x)w(x-\epsilon)\over    w(x+m) w(x-m-\epsilon) }    =   w(x) P(x) f_m(x) \label{eqw}
   \ee
   with $f_m(x)$ an unknown function with only $m$ dependent zeros and poles.
Indeed at large $m$ this equation  reduces to that of the pure $SU(N)$  gauge theory case after rescaling $ q m^{2N} \to q$ 
and setting $f_m \to 1$.   This implies in particular that $f_m(x)$ has only m dependent zeros and poles as claimed.

   Taking the log of both sides of (\ref{eqw}), multiplying by ${1\over z-x}$
   and integrating over a contour $\gamma$ including all the zeros and poles of $w(x)$, but not z,
   we can convert this equation into an integral equation for $w(x)$
    \be
  \framebox[1.15\width ][c]{$  \ln w(x) =- \ln P(x)
  +\int_\gamma {dz\over 2\pi  i(x-z)
  }\,
 \ln  \left( 1-q {w(z)w(z-\epsilon)\over  w(z+m)
 w(z-m-\epsilon)} \right)
  $}   \label{sadadj2}
   \ee
    where we use the fact that $f_m(x)$ has no zeros or poles inside $\gamma$ in order to discard
    its contribution to the integral in the right hand side.
    This equation can be easily solved perturbatively in $q$,
   writing $w(x)=\sum_{k=0}^\infty w_k(x) q^k$ and solving recursively for $w_k(x)$.
  The integral reduces then to a sum over residues at $x=e_{u}+i \epsilon$ with $i=0,1,..$.
For the first few terms one finds \footnote{In finding $w_1(x)$  we deform the contour $\gamma$ to
   include $z=x$ and $z=\infty$ and compute the corresponding residues. }
 \bea
 w_0(x) &=& {1\over P(x)}\nn\\
 {w_1(x)\over w_0(x)} &=& 1- {  P(x+m)P(x-m-\epsilon)
  \over  P(x) P(x-\epsilon)
}
 \eea
The chiral correlators and the prepotential are given in terms of this function
 via (\ref{trphij}) and (\ref{qdqf}). In particular one finds
 \bea
 a_u &=&  - \sum_{i=0}^\infty {\rm Res}_{z=e_u+i\epsilon}\,  z\,\partial_z  \ln
w(z)\label{aeadj}\\
&=& e_u -{q  \over P'(e_u)} \,\sum_{\kappa=\pm} {P(e_u- \kappa\,  m) P(e_u+\kappa\, m+\kappa \, \epsilon)
  \over  P(e_u+\kappa \, \epsilon)}
+ O(q^2)\nn
 \eea

\subsection{Quantizing the curve}

In this section we consider the non commutative version of the SW curve
for the case of the $SU(N)$  gauge theory with adjoint matter. The SW curve in this case is still
given by a polynomial of order $N$ in $x$ but now the coefficients of the polynomial are
given by modular functions involving  infinite powers of $e^{\pm z}$.
On the other hand, as we have seen in the last section, the deformed SW
differential  is found as a solution of an integral rather than a difference equation.
The connection between  the two descriptions is far from obvious and unlike the case
of matter in the fundamental representation we are not able to build the complete dictionary.
 Still, we will show that the non commutative version of the
 SW curve again captures the physics of the $\epsilon$ deformation.
 In particular we will show that the  prepotential of the SU(2) theory with adjoint matter
 is reproduced by the difference equation following from promoting the
 coordinates $(x,z)$ to non commutative variables.
   A more complete analysis of this case deserves further investigations.

The SW curve for  the $SU(N)$  gauge theory  with adjoint matter can be written in the
compact form \cite{D'Hoker:1997ha} \footnote{
Explicitly
$W(x,e^z)=\sum_{n=0}^N {1\over n!} {  \vartheta^{(n)}_1(z)\over  \vartheta_1(z)} (m\,\partial_x)^n P(x)$,
with $\vartheta^{(n)}_1(z)=\partial_z^n\, \vartheta_1(z)$.}
\be
W(x,e^z)= \vartheta_1\left( z-m  \partial_x  \right) P(x) =0
\ee
with
 \be
\vartheta_1(z)=\sum_{r\in \Z+\ft12}  q^{{r^2\over 2}}\, e^{(z-\pi
i)   r} \ee and $P(x)$ a polynomial of order $N$. A
non commutative version of this curve can be written by taking
$z\to \epsilon \partial_x $ and introducing the wave function
$\Psi(x)$ annihilated by the curve, i.e. \be \Psi(x) \,
\vartheta_1\left( \epsilon \overleftarrow{\partial}_x -m
\overrightarrow{\partial}_x \right)P(x) =0 \ee with the arrows
indicating whether the derivatives act on $P(x)$ or on $\Psi(x)$.
Expanding on $m$ one finds
 \be
\framebox[1.15\width ][c]{$ \sum_{r\in \Z+\ft12 }\sum_{n=0}^N
e^{\pi i r} \, q^{{r^2\over 2}}\,
 \Psi(x-r \epsilon)\,{r^n\over n!}\,(m\partial_x)^n P(x)\,=0   $}    \label{diffeqadj}
 \ee
 Dividing by $\Psi(x+\ft{\epsilon}{2})$ this equation can be rewritten entirely in terms of the
  ratio
  \be
  y(x)={\Psi(x-\ft{\epsilon}{2})\over  \Psi(x+\ft{\epsilon}{2}) }
  \ee
 The resulting difference equation can be easily solved perturbatively order by order in $q$,
 writing $y(x)=\sum_k y_k(x) q^k$ and solving for $y_k(x)$.

\subsubsection*{Example:  $SU(2)$ gauge theory}

 We illustrate our analysis in the case of the $SU(2)$  gauge theory.
 From (\ref{diffeqadj}), taking $P(x)=x^2-u$ and diving by
  $\Psi(x+\ft{\epsilon}{2})$ one finds for the first few orders
 \bea
  0 &=& \left[ (x-\ft{m}{2})^2-u \right] - y(x) \left[ (x+\ft{m}{2})^2-u \right]\\
  && +q \left(
  y(x) y(x-\epsilon)\left[ (x+\ft{3m}{2})^2-u\right] -{\left[ (x-\ft{3m}{2})^2-u \right]
  \over y(x+\epsilon)} \right) +O(q^3)\nn
 \eea
  Writing $y(x)=\sum_k y_k(x) q^k$ and solving for $y_k(x)$ gives
  \bea
  y_0(x) &=& {(x-\ft{m}{2})^2-u\over (x+\ft{m}{2})^2-u}\\
  {y_1(x)\over y_0(x)} &=&- {x m (m^2-\epsilon^2)
  \left[8(x^2-u) (3 m^2-6 u-2 x^2+2 \epsilon^2)+3 m^4-12 m^2 \epsilon^2
  \right] \over  4\prod_{l=0}^1 \left[ (x+\ft{m}{2}-l \epsilon)^2-u \right]
  \left[ (x-\ft{m}{2}+l \epsilon)^2-u \right] } \nn
  \eea
and so on. $y(x)$ has four sets of poles.
The analog of the periods of the ``standard'' SW curve are given by 
  \bea
a &=& -\int_{\gamma_u} {dx \over 2 \pi i}  (x+\ft{m}{2}) \partial_x \ln y(x)
=-\sum_{k=0}^\infty {\rm Res}_{x=\sqrt{u} -\ft{m}{2}+k \epsilon}
\left[(x+\ft{m}{2}) \partial_x \ln y(x) \right]\nn\\
&=& \sqrt{u}- q\, {m^2(m+\epsilon)^2\over \sqrt{u} (4\, u-\epsilon^2)}+O(q^2) \label{aa}
\eea
 where $\gamma_u$ is a contour surrounding one of the four sets of poles. 
The four choices are equivalent, for definitiveness we take
 the poles at $\sqrt{u}- \ft{m}{2}+k \epsilon$.
 (\ref{aa}) perfectly matches\footnote{We checked the agreement till order
 $q^3$.} the result (\ref{aeadj})  coming from the integral equation for the deformed
 differential after taking $e_u=\pm e$ and
 identifying
  \be
 \langle {\rm tr} \Phi^2 \rangle= 2e^2 =2u-4m(m+\epsilon) \sum_{k|d} \, d \,q^k \label{phi2qk}
 \ee
  with the sum over the divisors $d$ of $k$. 
(\ref{phi2qk}) is the $\epsilon$ deformed version of the result in \cite{Fucito:2005wc}.
This provides a rather convincing evidence that the non commutative
  deformation of the SW curve captures the physics of the $\epsilon$ corrections of the
  ${\cal N}=2^*$ theory.
A complete proof of this equivalence would be very welcome.

 \vskip 1cm \noindent {\large {\bf Acknowledgments}} \vskip 0.2cm

We thank D. Fioravanti, A. Okounkov and O.Ragnisco for useful discussions and 
N. Dorey for correspondence.
This work was partially supported by the ERC Advanced Grant n.226455 {\it ``Superfields''}, by the
Italian MIUR-PRIN contract 20075ATT78, by the NATO grant
PST.CLG.978785, European Commission FP7 Programme 
Marie Curie Grant Agreement PIIF2-GA-2008-221571 and Institutional Partnership grant of the Humboldt  Foundation of Germany.

\noindent \vskip 1cm

\begin{appendix}

\section{Counting functions: The toolkit}
\label{stoolkit}
 In this appendix we collect some properties of the counting functions
 ${\cal Y}(x)$ and $w(x)$ appearing in the text. These functions can be written
 in terms of the entire functions $Y$, $Y_0$ defined as
    \bea
   Y(z)&=& e^{{z\over \epsilon} \sum_u \psi\left({a_u\over \epsilon} \right)        } \prod_{vj}  \left( 1-{z\over x_{vj}} \right) e^{  {z\over x_{vj}^0} } \nn\\
   Y_0(z)&=&e^{{z\over \epsilon} \sum_u \psi\left({a_u\over \epsilon} \right)        } \prod_{vj}  \left( 1-{z\over x^0_{vj}} \right) e^{  {z\over x_{vj}^0} }
   \eea
   with
   \be
   x^0_{ui}=a_{ui}+(i-1) \epsilon
   \ee
   and $\psi(z)={\Gamma'(z)\over \Gamma(z)}$.
     Functions $Y(z)$and $Y_0(z)$ are holomorphic with zeros in $x_{ui}$ and $x^0_{ui}$ respectively.
They generalize the familiar infinite products entering in the definition of the $\Gamma$ function
 \bea
&& \prod_{n=1}^\infty \left(1+{z\over n} \right) \, e^{-{z\over n}}={ e^{-\gamma z}\over  \Gamma(z+1)} \nn\\
  &&\prod_{n=1}^\infty \left(1+{z\over n+a} \right) \, e^{-{z\over n+a}} = {\Gamma(1+a)\over \Gamma(1+a+z)}
  e^{\psi(a+1)z}   \label{yy0}
  \eea
  with $\gamma$ the Euler-Mascheroni constant.
 The second equation follows from the first one after using
   \be
   \psi(a+1)=-\gamma +\sum_{n=1}^\infty \left( {1\over n}-{1\over n+a}\right)
   \ee
 The second equation in (\ref{yy0}) gives an alternative writing for the $Y_0$ function
 as
\be
Y_0(z) =\prod_v  {\Gamma\left({a_v\over \epsilon}\right)   \over
 \Gamma\left({a_v-z\over \epsilon}\right) }
\ee
and implies
\be
  {Y_0(z)\over Y_0(z-\epsilon)}={   (-)^N P_0(z) \over   \epsilon^N }   \label{yoyo}
  \ee
Then one finds
\bea
{\cal Y}(x) &=& c\, { Y(x)\over Y_0(x)}    \qquad ~~~~~~~~  c=\prod_{ui} {x_{ui}\over x^0_{ui}} \nn\\
w(x) &=&{{\cal Y}(x-\epsilon)\over {\cal Y}(x) P_0(x) }={(-)^N\over \epsilon^N}  {Y(x-\epsilon)\over
Y(x)}
\eea
where the last equation follows from (\ref{yoyo}).
The saddle point equations for the case of fundamental and adjoint matter
  can be written in terms of $Y,Y_0$ in the form
\be
0=   1-{q\, M(x_{ui})\over \epsilon^{2N} } \,{ Y(x_{ui}-\epsilon )\over Y(x_{ui}+\epsilon)}
\ee
\be
  0 =    1-q \,      { Y(x_{ui}-\epsilon) Y(x_{ui}-m) Y(x_{ui}+m+\epsilon)\over
  Y(x_{ui}+\epsilon) Y(x_{ui}+m) Y(x_{ui}-m-\epsilon)  }
\ee
 Alternatively one can define a finite size version of the functions ${\cal Y}(x)$, $w(x)$
 \bea
{\cal Y}_L(x) &=&\prod_{u=1}^N \prod_{i=1}^L {x_{ui}-x \over x_{ui}^0-x }  \label{yw2}\\
w_L(x) &=&{{\cal Y}_L(x-\epsilon)\over {\cal Y}_L(x) P_0(x) }
= {1\over P_0(x-L \epsilon )}
 \prod_{u=1}^N \prod_{i=1}^L {x-x_{ui}-\epsilon \over x-x_{ui} } \nn
\eea
where in the last line we made use of the identity
\be
\prod_{u=1}^N \prod_{i=1}^L {x-x^0_{ui}\over x-x_{ui}^0-\epsilon}={P_0(x)\over
 P_0(x-L \epsilon )}
\ee
 The last equation in (\ref{yw2}) gives an alternative definition for the $w(x)$ counting function
\be
w(x) =\lim_{L\to \infty} {1\over (-L\epsilon)^N} \prod_{u=1}^N \prod_{i=1}^L {x-x_{ui}-\epsilon\over x-x_{ui} }
\ee

\section{The effective Hamiltonian from the profile function}
\label{profileappend}
In this appendix we present an alternative derivation of the saddle point equations starting
from the partition function written as an integral over the profile function $f(x)$ describing the
Young tableaux. The function $f(x)$, as it was the case for $\rho(x)$,  is completely determined in terms of the $x_{ui}$
and  the use of one or the other  is just a matter of tastes.

In \cite{Nekrasov:2003rj} it was showed that the $\epsilon$ deformed
partition function for the pure $SU(N)$ gauge theory can be written as 
\be
Z(\e_1,\e_2;\Lambda)=\int Df\, e^{-{1\over \epsilon_1 \epsilon_2} {\cal
H}_{\e_1,\e_2} (f)} 
\ee
with 
\be 
{\cal H}_{\epsilon_1, \epsilon_2}
(f)=\frac{\e_1\e_2}{4}\int dxdy
f''(x)f''(y)\g_{\e_1,\e_2}(x-y;\Lambda) \label{hh} 
\ee 
The function $\g_{\e_1,\e_2}(x;\Lambda)$ is defined as
\bea
 \g_{\e_1,\e_2}(x;\Lambda)=\frac{d}{ds} \Big|_{s=0}\frac{\Lambda^s}{\G(s)}\int_0^{\infty} \frac{dt}{t}
 \frac{t^s e^{-tx}}{(e^{t\e_1}-1)(e^{t\e_2}-1)}   \label{gamma}
\eea
and $f(x)$ is a piecewise constant function 
\bea f(x)
&=&\sum_u |x-a_u|+\sum_{ui} \left( |x-x_{ui}-\epsilon_1|-|x-x_{ui}-\epsilon_1-\epsilon_2| \right.\nn\\
&& ~~~~~~~~~~~~~~~~~~\left. -|x-x^0_{ui}|+|x-x^0_{ui}-\epsilon_2|\right)
\label{ff}
\eea
describing the profile of the Young tableaux  set $\{ Y_u\} $ (see figure 1). The profile function is specified by the
set $\{ x_{ui} \}$, with $x_{ui}$ describing the height of the $i^{\rm th}$ column
 of the Young tableau $Y_u$ centered at $a_u$.
 
\begin{figure}[t!]
\begin{center}
\includegraphics[scale=0.32]{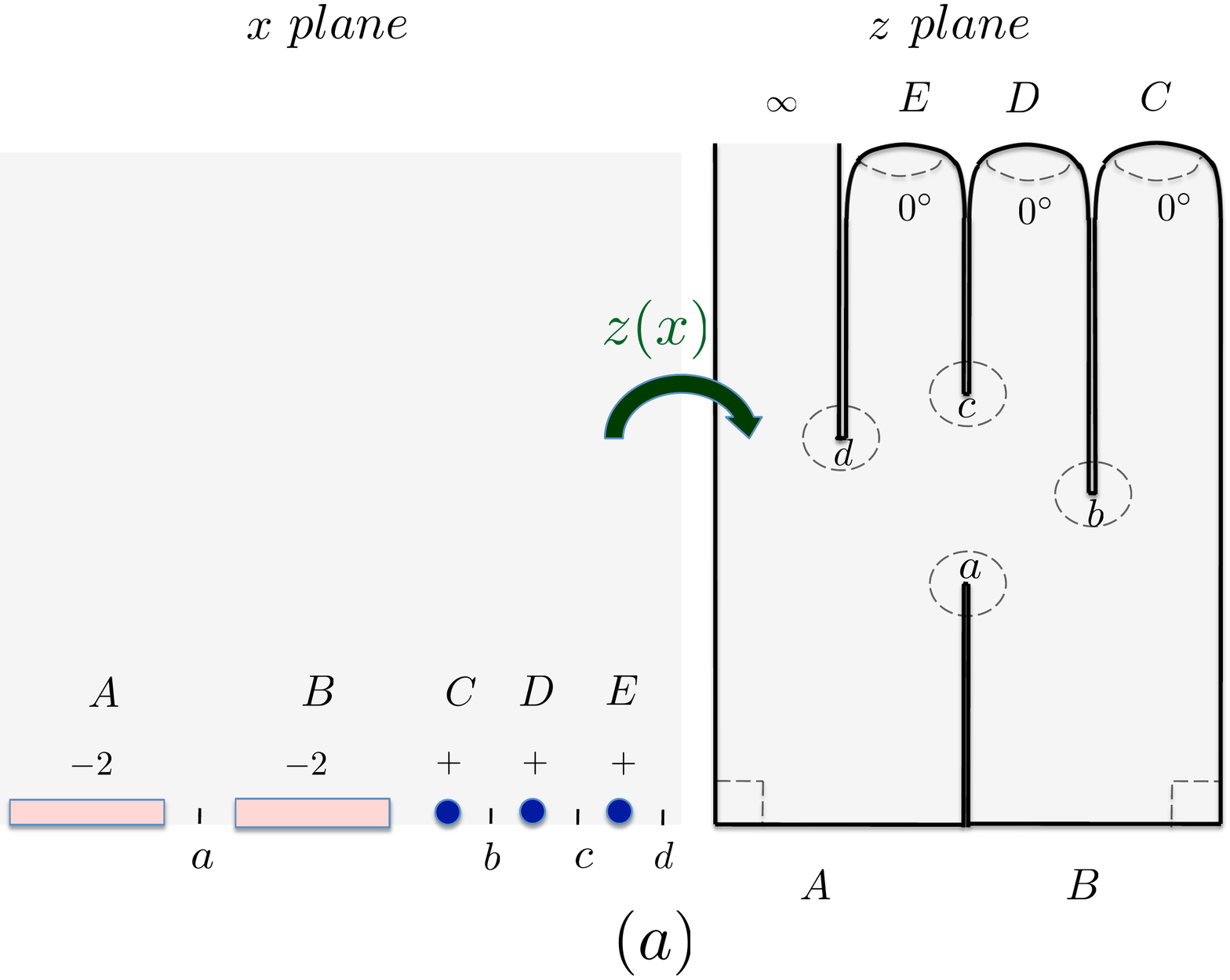}
\includegraphics[scale=0.32]{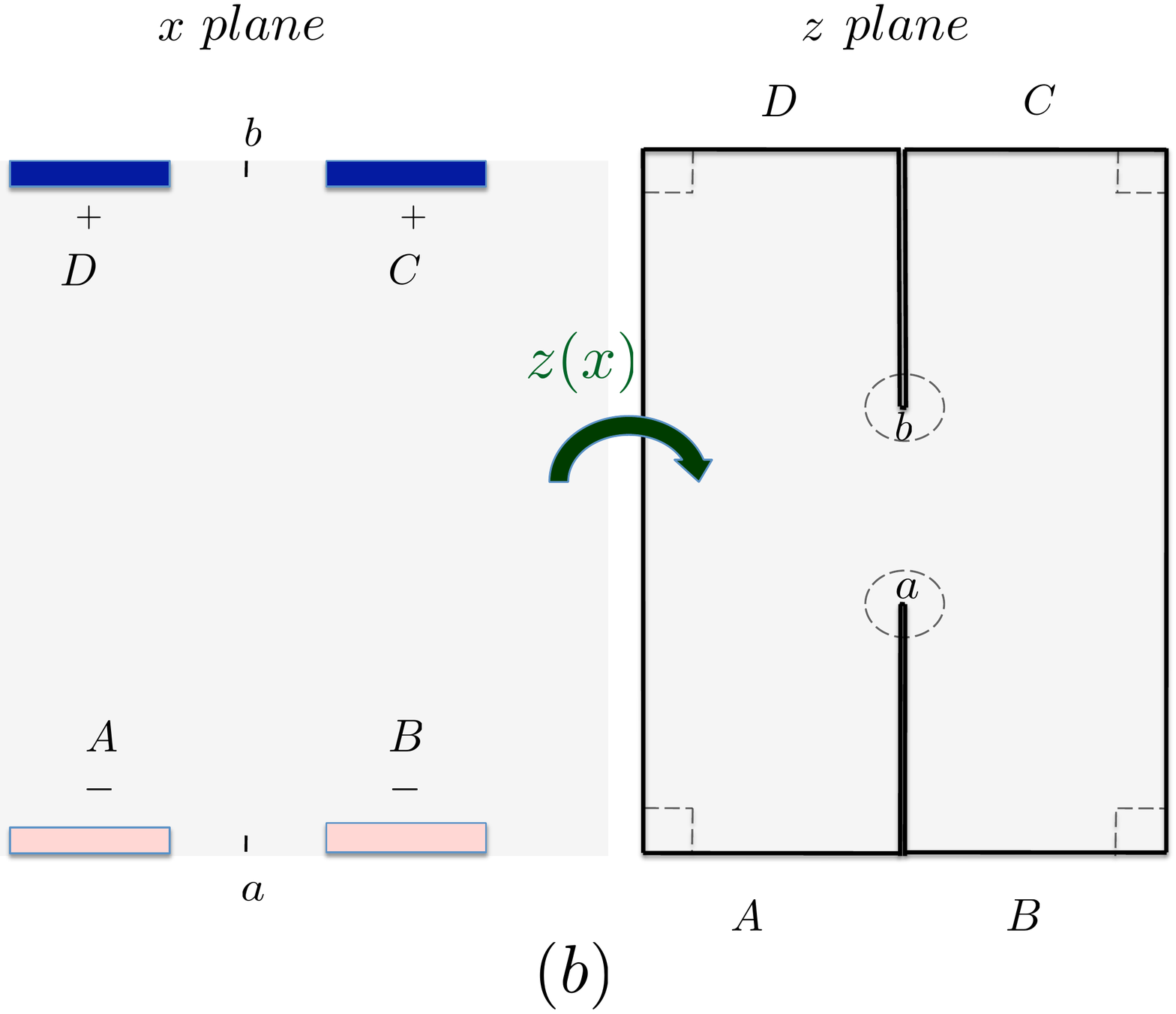}
\caption{\footnotesize Electrostatic problem associated to the $\epsilon_1,\epsilon_2 \to 0$ 
limit of $SU(2)$ gauge theory with  $N_f=3$ 
fundamental matter (a) or  one Adjoint hypermultiplet (b). }
\label{figelect2d}
\end{center}
\end{figure}

 The saddle point equation follows from extremizing  ${\cal
H}_{\e_1,\e_2} (f)$ with respect to $f$ and reads
\be 
   \int dy
f''(y) \left[ \g''_{\e_1,\e_2}(x-y;\Lambda)+ \g''_{\e_1,\e_2}(y-x;\Lambda)\right]  =0 \label{shh} \ee 
  where the variation is taken over piecewise functions of type (\ref{ff}). In the limit
 $\epsilon_1 \epsilon_2 \to 0$, the profile function $f(x)$  becomes a smooth function and the
 saddle point equation (\ref{shh}) reduces to\footnote{Here we use 
 $ \lim_{\epsilon_\ell \to 0} \epsilon_1 \epsilon_2 \g_{\epsilon}(x;\Lambda) \approx
  -\frac{x^2}{2}\log\Big(\frac{x}{\Lambda}\Big)+\frac{3}{4}x^2$. }
  \be 
 \int dy
f''(y) \log \Big| {x-y\over \Lambda } \Big| =0  \qquad {\rm for} \quad x\in \Sigma_u 
\label{elecpro}\ee 
 where $\Sigma_u$ is an interval around $a_u$ where $f''(x)$ is non zero. Notice 
that the right hand side of this equation can be interpreted as the two dimensional 
electrostatic potential generated by a continuous charge distribution $f''(x)$  along $\Sigma_u$.
The saddle point equation becomes then the condition that the potential is constant along 
$\Sigma_u$, i.e. $\Sigma_u$ can be thought as  a metallic plate\footnote{The extreme points
of $\Sigma_u$ are determined by solving the electrostatic problem.}. 
The electrostatic problems are illustrated in figure 2 for the case 
of $SU(2)$ gauge theory with $N_f=3$ and adjoint matter. The map $z(x)$ sends the real line to a domain in the complex plane
in such a way that its imaginary is constant along the plates, while its real part  is
constant on the path with no electric charges.
This domain is shown on the right hand side of figures (a) and (b). The explicit form of $z(x)$ is given by the
Christoffel formula
\be
z(x)=-\log w(x)=\int^x \prod_{a} (y-y_a)^{\varphi_a/\pi-1} dy
\ee
This map is specified by the points $y_a$ and angles $\varphi_a$ of  the polygon in the right hand side
 of figures (a) and (b). The integrand $dz(x)$ in this formula
is related to the SW differential.

 In this paper we study the saddle point equations for $\e_1\to 0$ with $\e_2=\epsilon$ finite. 
 This can be thought  as a discretization of the electrostatic problem we have just exposed where
 the metallic plates split into an infinite number of plus-minus dipoles  
 and the right hand side polygons in the figure  are replaced by polygons with only vertical lines (all charges are point like).
 The Hamiltonian become 
\be
{\cal H}_\epsilon(f) \equiv \lim_{\epsilon_1\to 0} {\cal H}_{\e_1,\e_2} (f)
\ee
In this limit  the $\gamma$ function reduces to
\bea
\gamma_{\e}(x;\Lambda) &=& \lim_{\e_1\to 0} \epsilon_1 \g_{\e_1,\e_2}(x;\Lambda)  \frac{d}{ds} \Big|_{s=0}\frac{\Lambda^s}{\G(s)} \int_0^{\infty} dt\,
 \frac{t^{s-2} e^{-tx}}{ (e^{t\epsilon}-1)} \nn\\
&=& \epsilon \frac{d}{ds} \left[  \frac{\Lambda^s}{ (s-1) \epsilon^{s}  } \zeta(s-1,1+\ft{x}{\epsilon})\right]_{s=0} \label{gammae}
\eea
with $\zeta(s,x)$ the Hurwitz zeta function. The propagator $\g''(x)$ entering in the saddle point
equations becomes\footnote{Here we use the following Zeta function identities:
  $\zeta(-1,x)=\ft{1}{12}+\ft{x}{2}(1-x) $, $\partial_z \partial_x \zeta(z,x)\Big|_{z=-1}
  =\log \Gamma(x)+x-\ft12(\log 2\pi+1)$.  }
  \be
  \gamma_\epsilon''(x;\Lambda)  {d\over dx} \log \Gamma(1+\ft{x}{\epsilon})+{1\over \epsilon} \log {\Lambda\over \epsilon} \label{g2}
  \ee
    In addition the profile function (\ref{ff}) reduces to
\bea
f''(x)&=& 2\sum_u \delta(x-a_u)+   2\sum_{ui} \left[ \delta(x-x_{ui})-\delta(x-x_{ui}-\epsilon)\right.\nn\\
&&\left. -
\delta(x-x^0_{ui})+\delta(x-x^0_{ui}-\epsilon)
\right] \nn\\
&=&  2\sum_{ui} \left[ \delta(x-x_{ui})-\delta(x-x_{ui}-\epsilon)\right]
\label{fdd}
\eea
  where in the second line we used the relation $x_{ui}^0+\epsilon=x^0_{u,i+1}$ to cancel $x_{ui}^0$-dependent terms against  the $ \delta(x-a_u)$ contributions.
Plugging  (\ref{g2}) and  (\ref{fdd}) into (\ref{shh}) 
one finds
\be
 {d\over dx} \log
\prod_{vj} {(x-x_{vj}+\epsilon) 
\over (x-x_{vj}-\epsilon )  }=0 \qquad {\rm for} \quad x=x_{ui}
\ee
 in agreement with (\ref{sadfund}) after taking $M(x)=1$. 
 
 We remark that $Z$ contains both perturbative and non-perturbative contributions to the
 instanton partition function. The perturbative part can be easily extracted
 by taking $q=0$, i.e.
 $x_{ui}=x_{ui}^0$. In this limit one finds $f''(x)=2\sum_u  \delta(x-a_u) $
 leading to
 \be
Z_{\rm pert} =  e^{-{1\over \epsilon_1}   
\sum_{u,v} \g_{\epsilon}(a_u-a_v;\Lambda)  } \label{hpert}
 \ee 
In the case of matter in the
fundamental and adjoint representations \cite{Nekrasov:2003rj,Nekrasov:2004vw,Shadchin:2004yx}, ${\cal H}_{\e_1,e_2}$ can be treated in a similar way.  
In these cases the additional contributions to the Hamiltonian  
are given by 
 \bea
 {\cal H}_{\rm fund}(f) &=& -\ft{\epsilon_1 \epsilon_2}{2} \sum_a \int dx f''(x) \gamma_{\e_1,e_2} (x+m_a;\Lambda)\\
 {\cal H}_{\rm adj}(f) &=& -\frac{\e_1 \e_2}{4}\int dxdy f''(x)f''(y)\g_{\e_1,\e_2}(x-y+m;\Lambda)
+\ft{\log q}{4} \int dx x^2 f''(x)\nn
\eea
Following the same manipulations as before one can easily reproduces the saddle point equations
(\ref{sadfund}) and (\ref{sadadj}). The matter contributions to the perturbative partition function 
read 
\bea
 Z_{\rm fund,pert}(f) &=& e^{{1\over \epsilon_1} \sum_{a,u} 
 \gamma_{\epsilon} (a_u+m_a;\Lambda)} \nn\\
Z_{\rm adj,pert}(f) &=& e^{{1\over \epsilon_1} \left( \sum_{u,v} \g_{\epsilon}(a_u-a_v+m;\Lambda)
-\ft{\log q}{2} \sum_u a_u^2\right) } 
\eea


\section{Testing the deformed SW differentials}

In this appendix we test the prepotential following from the deformed SW differential
against a direct computation based on the partition function. We restrict ourselves to
$k=1$.

\subsection{SU(N) plus fundamental matter}

The $k=1$ SW prepotential in the limit $\epsilon_1 \to 0$ is given by
\bea
{\cal F}_1 &=& -\lim_{\epsilon_1 \to 0} \epsilon_1 \epsilon_2 Z_1
=-\epsilon \int_{\R} {dx\over 2 \pi i} {M(x) \over P_0(x+\epsilon ) P_0(x )}\nn\\
&=& -\epsilon \sum_u  {M(a_u) \over P_0(a_u+\epsilon ) P_0'(a_u )} \label{f1fund}
\eea
On the other hand using (\ref{qdqf}) and the results (\ref{aefund}) coming from the deformed SW curve
one finds
\bea
2 \sum_{k=1}^\infty  k {\cal F}_k q^k &=& \sum_u (e_u^2-a_u^2)\nn\\
&=& 2 q\,
\sum_u  {a_u\over P_0'(a_u)} \left[  { M(a_u)
  \over  P_0(a_u+\epsilon)}+   { M(a_u-\epsilon)
  \over  P_0(a_u-\epsilon)}  \right] +O(q^2) \nn\\
  &=&- 2 q \epsilon \sum_u {M(a_u) \over P_0(a_u+\epsilon ) P_0'(a_u )}+O(q^2)
\eea
in agreement with (\ref{f1fund}) for $k=1$. In deriving the last equation we use the identities
\bea
&& \sum_u    \left[  { M(a_u)
  \over  P_0'(a_u) P_0(a_u+\epsilon)}+   { M(a_u-\epsilon)
  \over  P_0'(a_u) P_0(a_u-\epsilon)}  \right]=0\nn\\
&& \sum_u    \left[  { (a_u+\epsilon) M(a_u)
  \over  P_0'(a_u) P_0(a_u+\epsilon)}+   { a_u M(a_u-\epsilon)
  \over  P_0'(a_u) P_0(a_u-\epsilon)}  \right]=0
\eea
that follow from the vanishing of the contour integral around infinity of the function
${(x+\epsilon)^a M(x) \over P_0(x+\epsilon ) P_0(x )}$
with $a=0,1$.

\subsection{SU(N) plus adjoint matter}

The $k=1$ SW prepotential in the limit $\epsilon_1 \to 0$ is given by
\bea
{\cal F}_1 &=& -\lim_{\epsilon_1 \to 0} \epsilon_1 \epsilon_2 Z_1
= -\epsilon
\int_{\R} {dx\over 2 \pi i} {P_0(x+m+\epsilon ) P_0(x-m ) \over P_0(x+\epsilon ) P_0(x )}\nn\\
&=& -\epsilon \sum_u  {P_0(a_u+m+\epsilon ) P_0(a_u-m ) \over P_0(a_u+\epsilon ) P_0'(a_u )} \label{f1adj}
\eea
On the other hand using (\ref{qdqf}) and the results (\ref{aeadj}) coming from the deformed SW curve
one finds
\bea
&&2 \sum_{k=1}^\infty  k {\cal F}_k q^k = \sum_u (e_u^2-a_u^2)\nn\\
&&\quad = 2 q\,
\sum_u  {  a_u \over P'(a_u)} \,\sum_{\kappa=\pm} {P(a_u- \kappa\,  m) P(a_u+\kappa\, m+\kappa \, \epsilon)
  \over  P(a_u+\kappa \, \epsilon)} +O(q^2) \nn\\
  &&\quad = -2 q \epsilon \sum_u  {P_0(a_u+m+\epsilon ) P_0(a_u-m ) \over P_0(a_u+\epsilon ) P_0'(a_u )}+O(q^2)
\eea
in agreement with (\ref{f1adj}) for $k=1$. In deriving the last equations we follow similar manipulations
of the contour integral as in the case of fundamental matter.

\section{A TBA like equation}

 In  \cite{Nekrasov:2009rc} an intriguing correspondence between the dynamics of ${\cal N}=2$ gauge theories on 
the $\Omega_{\epsilon_1,\epsilon_2}$ deformed background and quantum integrable systems was proposed. 
According to this proposal the deformed prepotential ${\cal F}(\epsilon,a_u)$
 of the gauge theory  is identified with the Yang Yang function of the quantum integrable model
 with $\epsilon$ playing the role of  the Planck  constant.
 Finally the eigenvalues of the integrable Hamiltonians ${\cal H}_J$ are given by
 the chiral correlators $\langle {\rm tr} \Phi^J \rangle$ after quantizing $a_u$ by
 requiring ${\partial {\cal F}\over a_u}=n_u\in \Z$.

   In this section we show that the deformed SW equations we found in the text
   can be rewritten in a Thermodynamic Bethe Ansatz (TBA) like form of the type presented in
   \cite{Nekrasov:2009rc}.   For concreteness we focus on the ${\cal N}=2^*$ gauge theory
associated, according to \cite{Nekrasov:2009rc}, to the quantum elliptic Calogero-Moser system. The analysis
here can be easily adapted to the case of fundamental matter.
  We first rewrite (\ref{sadadj}) in the Bethe Ansatz like form
\be
\framebox[1.15\width ][c]{$  e^{-\varphi(x_{ui})}=1  $}   \label{bae}
\ee
with
\bea
-\varphi(x) &=& \ln \left(q {w(x) w(x+\epsilon)\over w(x+m+\epsilon) w(x-m)} \right) \label{varphi}
   \eea
Using (\ref{sadadj2}) we can rewrite  each $\ln w(x+A)$ in the right hand side of (\ref{varphi}) in an integral form. Collecting the various  pieces one finds  the TBA like equation
  \be
   \framebox[1.15\width ][c]{$ -\varphi(x)=\ln q Q(x)+\int_\gamma dy\, \tilde G(x-y) \ln (1-e^{-\varphi(y)}) $}
   \label{tba}
   \ee
with
 \bea
   \tilde G(x) &=& {d\over dx} \ln \left[ { x(x-\epsilon)\over (x+m) (x-m-\epsilon)}\right]\nn\\
   Q(x) &=& {P(x+m+\epsilon)P(x-m)\over P(x) P(x+\epsilon)}
    \label{g}
\eea
(\ref{tba}) is close to that found in \cite{Nekrasov:2009rc}. The two are related by the
 replacement  $\tilde G(x) \to G_s(x)$.  We don't fully understand the origin of this discrepancy.
  We should remark however that, despite the similarities, 
 (\ref{bae}) and (\ref{tba}) are very different from the standard Bethe Ansatz 
  and TBA equations  . The main difference being
that, unlike in the standard setting of a Bethe Ansatz, in the present case $e^{-\varphi}$
is not an unimodular phase and therefore $1-e^{-\varphi}$ has not only zeros at $x_{ui}+\epsilon$
but also poles at $x_{ui}$. This explains the appearance of $\tilde G(x)$ rather than $G_s(x)$
in our TBA like equation (\ref{tba}), in contrast with the expectations coming from integrable
systems where  the reflection matrix is a unimodular phase. 
 It would be nice to understand, may be along the lines of  
 \cite{Bombardelli:2009xz,Bombardelli:2009ns},  how TBA techniques extend to this case.

\end{appendix}

\providecommand{\href}[2]{#2}\begingroup\raggedright\endgroup

    \end{document}